\documentclass[english,fleqn,twoside]{article}
\usepackage[T1]{fontenc}
\usepackage[latin1]{inputenc}
\usepackage{graphicx}
\usepackage{amssymb}

\usepackage{color}
\definecolor{darkgreen}{rgb}{0,0.5,0}
\definecolor{purple}{rgb}{0.5,0,0.5}
\definecolor{nblue}{rgb}{0.0,0.0,0.50}
\definecolor{scarlet}{rgb}{1.0,0.2,0}

\usepackage[colorlinks=true, pdfstartview=FitV, linkcolor=purple, citecolor= purple, urlcolor=blue]{hyperref}

\makeatletter

 \usepackage{espcrc2}

\newcommand{\Eq}{Eq.~}

\newcommand{\lsim}{\mathrel{\rlap{\lower4pt\hbox{\hskip0pt$\sim$}} 
\raise1pt\hbox{$<$}}}           
\newcommand{\gsim}{\mathrel{\rlap{\lower4pt\hbox{\hskip0pt$\sim$}} 
\raise1pt\hbox{$>$}}}           

\usepackage{babel}
\makeatother
\begin{document}

\title{Schwinger functions and light-quark bound states, and sigma terms}

\author{A.\ H\"{o}ll,\address[Rostock]{Institut f\"ur Physik, Universit\"{a}t Rostock, D-18051 Rostock, Germany}
P.\ Maris,\address[Pitt]{Department of Physics and Astronomy, University of Pittsburgh, PA 15260, U.S.A.}
C.\,D.\ Roberts\,\addressmark[Rostock]$^{,}$\address[ANL]{Physics Division, Argonne National Laboratory, Argonne, IL 60439-4843, U.S.A.}
and S.\,V.\ Wright\,\addressmark[ANL]}

\begin{abstract}
We explore the viability of using solely spacelike information about a Schwinger function to extract properties of bound states.  In a concrete example it is not possible to determine properties of states with masses $\gsim 1.2\,$GeV.  
Modern Dyson-Schwinger equation methods supply a well-constrained tool that provide access to hadron masses and $\sigma$-terms.  We report values of the latter for a range of hadrons.  Of interest is analysis relating to a $u$,$d$ scalar meson, which is compatible with a picture of the lightest $0^{++}$ as a bound state of a dressed-quark and \mbox{-antiquark} supplemented by a material pion cloud.  
A constituent-quark $\sigma$-term is defined, which affords a means for assessing the flavour-dependence of dynamical chiral symmetry breaking.
\end{abstract}
\maketitle

\section{Introduction}
The pseudoscalar spectrum {[}$I^{G}(J^{PC})=1^{-}(0^{-+})${]} contains three states below 2 GeV: $\pi(140)$; $\pi(1300)$; and $\pi(1800)$.  The lightest has been much studied.  However, a comprehensive understanding of QCD requires an approach that admits the simultaneous study of the heavier pseudoscalars and, indeed, other systems.  An understanding of the hadron spectrum and its realisation within QCD is necessary in order to unravel the nature of the long-range force between light quarks.  

The lightest pseudoscalar meson is both a bound state of $u$- and $d$- quarks and the Goldstone mode in QCD associated with the dynamical breaking of chiral symmetry.  This is readily and clearly understood using the Dyson-Schwinger equations (DSEs) \cite{Maris:1997hd,Maris:1997tm}.  The DSEs have proven a particularly useful device for studying the spectrum and properties of light-quark systems.  Modern applications are reviewed in Refs.\ \cite{bastirev,reinhardrev,Maris:2003vk}.

Herein we present results obtained from studies of the inhomogeneous Bethe-Salpeter equations (BSEs) for the scalar and pseudoscalar vertices in QCD.  This is a practical means of mapping out the domain of applicability of the leading order term in a systematic and symmetry preserving DSE truncation scheme \cite{munczek,bender,mandarvertex}.  Furthermore, we explore the capacity of such studies to complement contemporary numerical simulations of lattice-regularised QCD.

\section{Bound states from spacelike data}
\label{sec:Numerical-Tools}
The BSE provides a Poincar\'e covariant tool with which to calculate the properties of bound states in quantum field theory.  The inhomogeneous equation for a pseudoscalar quark-antiquark vertex is\footnote{For simplicity we work with two degenerate flavours of quarks.  Hence, Pauli matrices are sufficient to represent the flavour structure.  We employ a Euclidean metric with the conventions described, e.g., in Sec.~2.1 of Ref.\,\protect\cite{bastirev}.}
\begin{eqnarray}
\nonumber \lefteqn{\left[\Gamma_5^j(k;P)\right]_{tu}= 
Z_{4}\gamma_{5}\,\frac{\tau^j}{2}  }\\
&&  +\int_{q}^{\Lambda}\left[\chi_5^j(q;P)\right]_{sr}K_{rs}^{tu}(q,k;P)\,,
\label{eq:DSE_inhomogeneous}
\end{eqnarray}
where $k$ is the relative and $P$ the total momentum of the constituents;
$r,\ldots,u$ represent colour, Dirac and flavour matrix indices;
\begin{equation}
\chi_5^j(q;P) = S(q_{+}) \Gamma_5^j(q;P) S(q_{-})\,,
\end{equation}
$q_{\pm}=q\pm P/2$; and $\int_{q}^{\Lambda}$ represents a Poincar\'{e}
invariant regularisation of the integral, with $\Lambda$ the regularisation
mass-scale \cite{Maris:1997hd,Maris:1997tm}.  In \Eq(\ref{eq:DSE_inhomogeneous}),
$K$ is the fully amputated and renormalised dressed-quark-antiquark scattering kernel and $S$ is the renormalised dressed-quark propagator. ($SSK$ is a renormalisation group invariant).  The dressed-quark propagator has the form 
\begin{eqnarray} 
 S(p)^{-1} 
& =& \frac{1}{Z(p^2,\zeta^2)}\left[ i\gamma\cdot p + M(p^2)\right] ,
\label{sinvp} 
\end{eqnarray} 
and is obtained as the solution of QCD's gap equation:
\begin{equation}
S(p)^{-1}=Z_{2}\left(i\gamma\cdot p+m_{\mathrm{bm}}\right)+\Sigma(p)\,,\label{eq:quark_prop}
\end{equation}
\begin{equation}
\Sigma(p) = Z_{1} \int_{q}^{\Lambda} g^{2} D_{\mu\nu}(p-q) \frac{\lambda^{a}}{2} \gamma_{\mu} S(q) \Gamma_{\nu}^{a}(q;p)\,,
\end{equation}
augmented by the renormalisation condition
\begin{equation}
\label{renormS} \left.S(p)^{-1}\right|_{p^2=\zeta^2} = i\gamma\cdot p +
m(\zeta)\,,
\end{equation}
where $m(\zeta)$ is the running current-quark mass at the renormalisation point $\zeta$.  These equations involve the quark-gluon-vertex, quark wave function and Lagrangian mass renormalisation constants, $Z_{1,2,4}(\zeta,\Lambda)$, each of which depends on the gauge parameter, the renormalisation point and the regularisation mass-scale.

\begin{figure}[t]
\includegraphics[clip,width=0.45\textwidth]{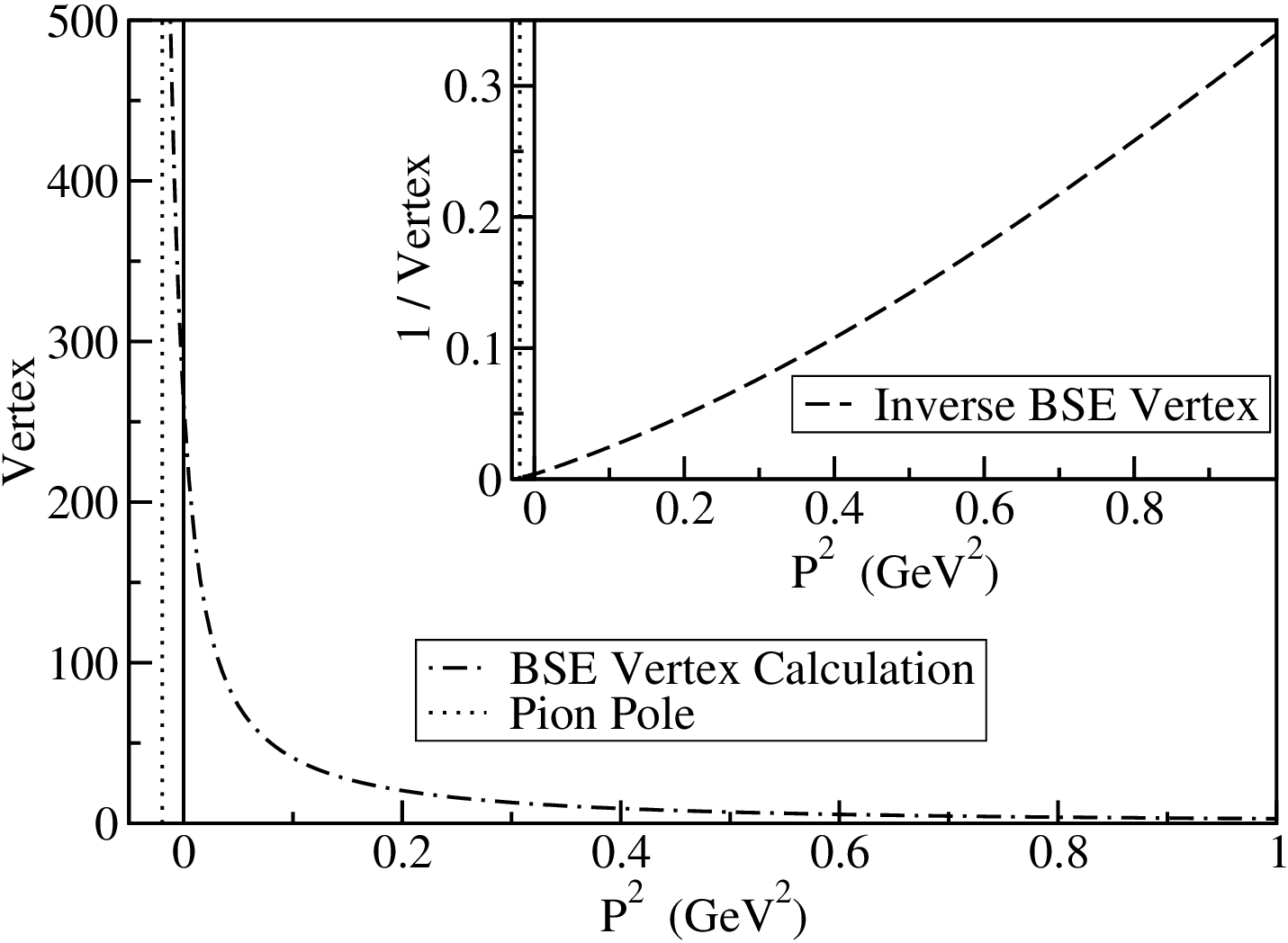}\vspace*{-4ex}

\caption{\label{fig:DSE_Inverted} Pseudoscalar amplitude $E_5(0;P^2)$ obtained by solving the inhomogeneous BSE, Eq.\,(\ref{eq:DSE_inhomogeneous}), using the renormalisation-group-improved rainbow-ladder truncation described in \protect\cite{mariscairns}.  The vertical dotted line indicates the position of the ground state $\pi$ mass pole.  The inset shows $1/E_5(0;P^2)$.\vspace*{-4ex}}
\end{figure}

The solution of Eq.\,(\ref{eq:DSE_inhomogeneous}) has the form 
\begin{eqnarray}
\nonumber
\lefteqn{i \Gamma_{5 }^j(k;P) = \frac{\tau^j}{2} \gamma_5
\left[ i E_5(k;P) + \gamma\cdot P \, F_5(k;P) \right.} \\
\nonumber & &
\left.+ \, \gamma\cdot k \,k\cdot P\, G_5(k;P)+
 \sigma_{\mu\nu}\,k_\mu P_\nu \,H_5(k;P) \right].\\
\label{genpvv}
\end{eqnarray}
This is the minimal and complete form required by Poincar\'e covariance.  The homogeneous pseudoscalar BSE is obtained from Eq.\,(\ref{eq:DSE_inhomogeneous}) merely by omitting the \emph{driving term}; viz., $Z_{4}\gamma_{5}\frac{\tau^j}{2} $.  The equation thus obtained defines an eigenvalue problem, with the bound state's mass-squared being the eigenvalue and its Bethe-Salpeter amplitude, the eigenvector.  As such, the equation only has solutions at isolated timelike values of $P^2$.  On the other hand, the solution of the inhomogeneous equation, Eq.\,(\ref{eq:DSE_inhomogeneous}), exists for all values of $P^2$, timelike and spacelike, with each bound state exhibited as a pole.  This is illustrated by Fig.\,\ref{fig:DSE_Inverted}, wherein the solution is seen to evolve smoothly with $P^2$ and the pole associated with the pseudoscalar ground state is abundantly clear. 

A numerical determination of the precise location of the first pole in $E_5(k^2=0;P^2)$ will generally be difficult.  The task becomes harder if one seeks to obtain the positions of excited states in addition.  It is for these reasons that the homogeneous equation is usually used.  However, if one is employing a framework that can only provide information which is equivalent to the form of this vertex at spacelike momenta, then a scheme must be devised that will yield the pole positions.  

One obvious alternative is to focus on $P_E(P^2)=1/E(k^2=0;P^2)$ and locate its zeros, and it is plain from Fig.\,\ref{fig:DSE_Inverted} that this approach can at least be successful for the ground state.  It is important to determine whether this (or another) approach can also be used in practice to determine some properties of excited states when information is available only for spacelike $P^2$.  

While the DSEs can be used to generate such information, and the analysis of that information is what brought us to this point, herein for this purpose we consider a simple model for an inhomogeneous vertex whose analytic structure is known precisely; namely, 
\begin{equation}
V(P^2)=b + \sum_{i=1}^M\frac{a_{i}}{P^{2}+m_{i}^{2}}\,, \label{eq:simple_model}
\end{equation}
where: for each $i$, $m_{i}$ is the bound state's mass and $a_{i}$ is the residue of the bound state pole in the vertex, which is related to the state's decay constant; and $b$ is a constant that represents the perturbative background that is necessarily present in the ultraviolet.  The particular parameter values we employ are listed in Table\,\ref{tab:Model_Param}.   This \textit{Ansatz} provides a data sample that captures the essential qualitative features of true DSE solutions for colour-singlet three-point Schwinger functions, such as that depicted in Fig.\,\ref{fig:DSE_Inverted}.

\begin{table}[t]
\begin{center}
\caption{\label{tab:Model_Param} Parameters characterising our vertex \textit{Ansatz}, Eq.\,(\protect\ref{eq:simple_model}).  They were chosen without prejudice, subject to the constraint in quantum field theory that residues of poles in a three-point Schwinger function must alternate in sign \cite{Holl:2004fr}, and ordered such that $m_i<m_{i+1}$.  We use $b=0.78$.  This is the calculated value of $Z_4(\zeta = 19\,{\rm GeV},\Lambda=200\,{\rm GeV})$ used to obtain the curves in Fig.\,\protect\ref{fig:DSE_Inverted}.}

\begin{tabular*}{0.45\textwidth}{
|c@{\extracolsep{0ptplus1fil}}|l@{\extracolsep{0ptplus1fil}}|l@{\extracolsep{0ptplus1fil}}|}\hline
$i$~ & Mass & Residue \\\hline
1& 0.14& ~4.23\\  
2& 1.06& -5.6 \\
3&1.72& ~3.82\\  
4&2.05 & -3.45 \\  
5&2.2& ~2.8 \\\hline
\end{tabular*}\vspace*{-4ex}
\end{center}
\end{table}

To proceed, we employ a diagonal Pad\'e approximant of order $N$ to analyse the data sample generated by Eq.\,(\ref{eq:simple_model}); viz., we use
\begin{equation}
f_{N}(P^{2}) = \frac{c_{0}+c_{1}P^{2}+\ldots+c_{N}P^{2N}} {1+c_{N+1}P^{2}+\ldots+c_{2N}P^{2N}}
\label{eq:pade}
\end{equation}
as a means by which to fit $1/V(P^2)$.  The known ultraviolet behaviour of the vertex requires that we use a diagonal approximant.  NB.\ A real-world data sample will exhibit logarithmic evolution beyond our renormalisation point.  No simple Pad\'e approximant can recover that.  However, this is not a problem in practical applications because the approximant is never applied on that domain.

In a confining theory it is likely that a colour-singlet three-point function exhibits a countable infinity of bound state poles.  Therefore no finite order approximant can be expected to recover all the information contained in that function.  Our vertex model exhibits $M$ bound state poles.  We expect that an approximant of order $N<M$ will at most provide reliable information about the first $N-1$ bound states, with the position and residue associated with the $N^{\rm th}$ pole providing impure information that represents a mixture of the remaining $M-(N-1)$ signals and the continuum.  We anticipate that this is the pattern of behaviour that will be observed with any rank-$N$ approximation to a true Schwinger function.  To explore this aspect of the procedure we studied the $N$-dependence of the Pad\'e fit.  

The domain of spacelike momenta for which information is available may also affect the reliability of bound state parameters extracted via the fitting procedure.  We analysed this possibility by fitting Eq.\,(\ref{eq:pade}) to our \textit{Ansatz} data on a domain $(0,P_{\mathrm{max}}^{2}]$, and studying the $P_{\mathrm{max}}^{2}$-dependence of the fit parameters.

In all cases we found that a Pad\'e approximant fitted to $1/V(P^2)$ can accurately recover the pole residues and locations associated with the ground and first excited states.  However, there is no cause for complacency. 

\begin{figure}[t]
\includegraphics[clip,width=0.45\textwidth]{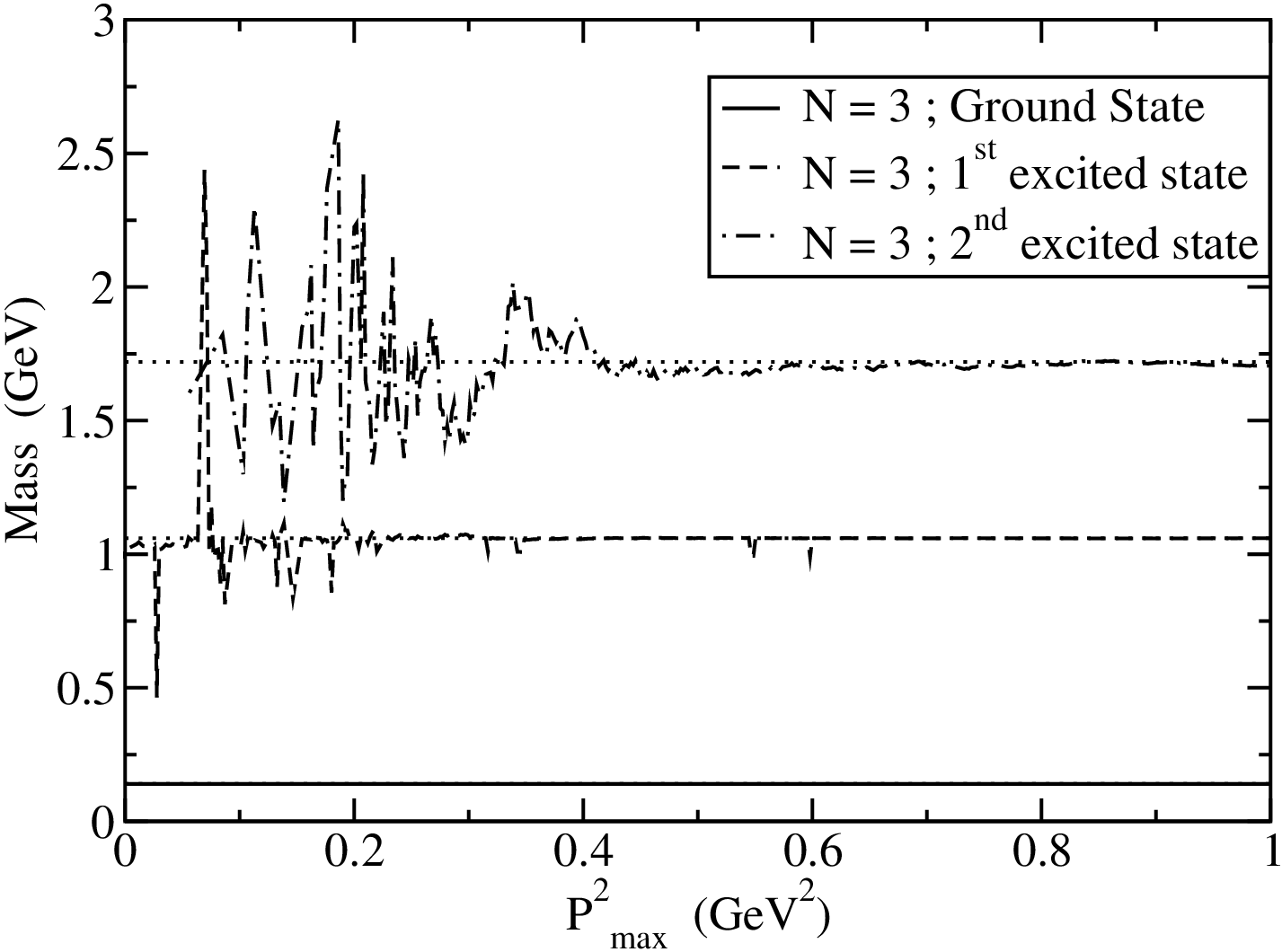}\vspace*{-4ex}

\caption{\label{fig:3pade_mass} Pole positions (mass values) obtained through a fit of Eq.\,(\protect\ref{eq:pade}) with $N=3$ to data for $1/V(P^2)$ generated from Eq.\,(\protect\ref{eq:simple_model}) with the parameters listed in Table \ref{tab:Model_Param}.  The coordinate $P_{\mathrm{max}}^{2}$ is described in the text. Horizontal dotted lines indicate the three lightest masses in Table \ref{tab:Model_Param}.  The ground state mass (solid line) obtained from the Pad\'e approximant lies exactly on top of the dotted line representing the true value.\vspace*{-4ex}}
\end{figure}

\begin{figure}[tb]
\includegraphics[clip,width=0.45\textwidth]{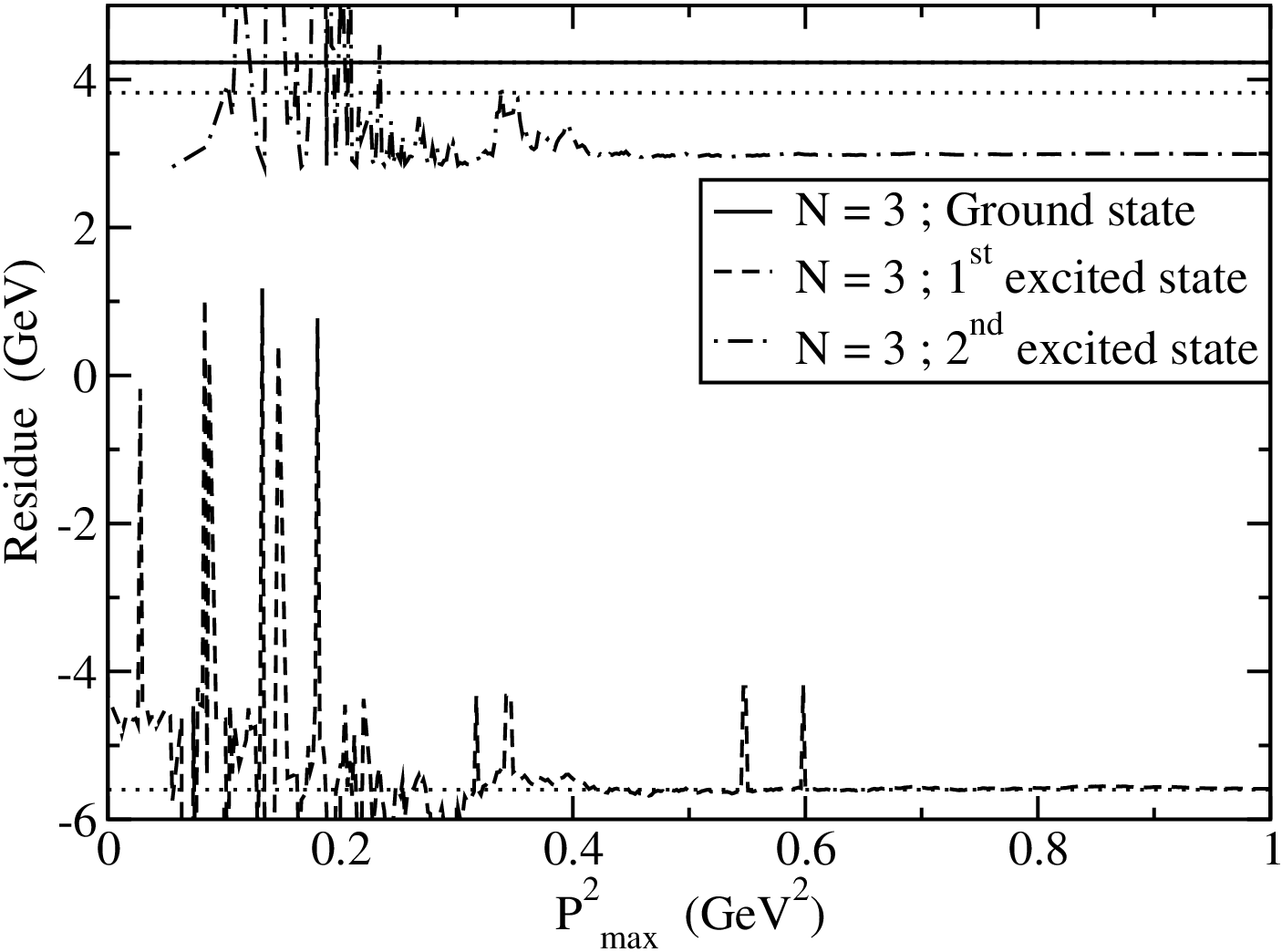}\vspace*{-4ex}

\caption{\label{fig:3pade_residue} Pole residues obtained through a fit of Eq.\,(\protect\ref{eq:pade}) with $N=3$ to data for $1/V(P^2)$ generated from Eq.\,(\protect\ref{eq:simple_model}) with the parameters in Table \ref{tab:Model_Param}.  Horizontal dotted lines indicate the residues associated with the three lightest masses in Table \ref{tab:Model_Param}.  The residue associated with the ground state (solid line) lies exactly atop the dotted line representing the true value.  The residue for the second pole exhibits a plateau at the correct (negative) value.  However, the plateau exhibited by the result for the residue of the third pole is wrong.\vspace*{-4ex}}
\end{figure}

In Fig.\,\ref{fig:3pade_mass} we exhibit the $P_{\mathrm{max}}^{2}$-dependence of the mass-parameters determined via a $N=3$ Pad\'e approximant.  Plateaux appear for three isolated zeros, which is the maximum number possible, and the masses these zeros define agree very well with the three lightest values in Table \ref{tab:Model_Param}.  This appears to suggest that the procedure has performed better than we anticipated.  However, that inference is seen to be false in Fig.\,\ref{fig:3pade_residue}, which depicts the $P_{\mathrm{max}}^{2}$-dependence of the pole residues.  While the results for $a_{1,2}$ are correct, the result inferred from the plateau for $a_3$ is incorrect.  It is important to appreciate that if we had not known the value of $a_3$ \textit{a priori}, then we would very likely have been misled by the appearance of a plateau and produced an erroneous \emph{prediction} from the fit to numerical data.  Plainly, a $N=3$ approximant can at most only provide reliable information for the first $M-N=2$ bound states.  

We explored this further and applied a $N=4$ approximant to the same model.  In this case we could still only extract reliable information for the first two poles.  We subsequently biased the fit procedure by \emph{hard-wiring} in Eq.\,(\ref{eq:pade}) the residue and position of the lightest pole masses.  This did not help.  The $N=4$ approximant could still not provide results that improved upon what we had already learnt with the $N=3$ approximant.  
 
The results described herein, and our continuing analysis and exploration of other methods, including those popular in contemporary simulations of lattice-regularised QCD \cite{latticecorrelator}, suggest that solely from spacelike data it is only ever possible to extract reliable information about bound states with masses which do not much exceed $1\,$GeV.  While these results are preliminary, they nevertheless provide sound reasons for caution.

\begin{figure}[tb]
\includegraphics[clip,width=0.45\textwidth]{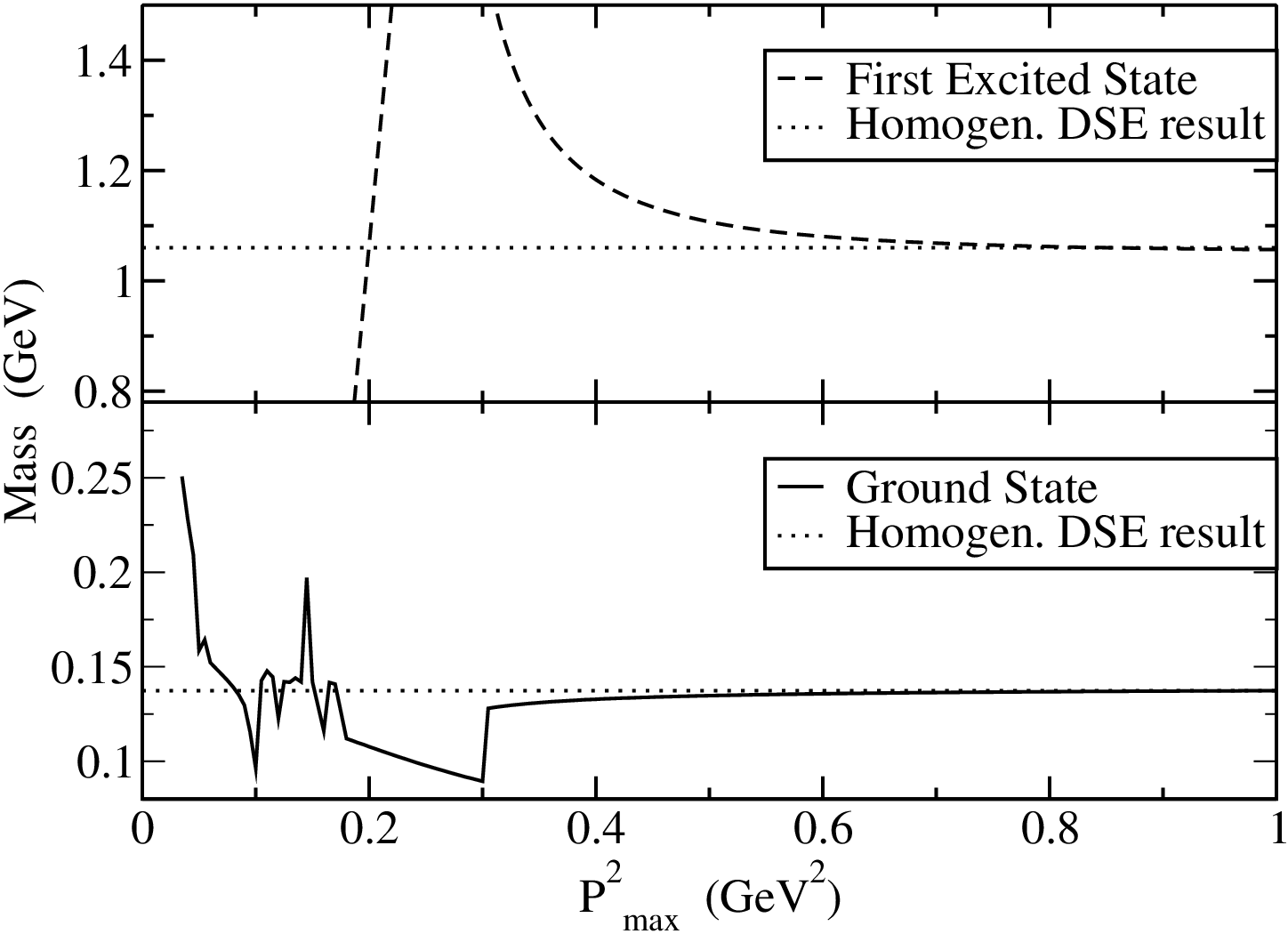}\vspace*{-4ex}

\caption{\label{fig:dse_masses} 
Pole positions (mass values) obtained through a fit of Eq.\,(\protect\ref{eq:pade}) with $N=3$ to the DSE result for $1/E_5(k^2=0,P^2)$ depicted in Fig.\,\protect\ref{fig:DSE_Inverted}.  Horizontal dotted lines indicate the masses obtained for the ground and first excited state via a direct solution of the homogeneous BSE \protect\cite{Holl:2004fr}.\vspace*{-4ex}}
\end{figure}

Following this background work on the viability \emph{in principle} of using spacelike data alone to extract bound state information, we applied the method to the true DSE-calculated pseudoscalar vertex that is in depicted Fig.\,\ref{fig:DSE_Inverted}.  The decay constants and masses for the ground state pseudoscalar and first radial excitation were obtained from the homogeneous BSE in Ref.\,\cite{Holl:2004fr}.  The comparison between these masses and those inferred from the Pad\'e approximant is presented in Fig.\,\ref{fig:dse_masses}.  As suggested by our background work, with perfect (effectively noiseless) spacelike data at hand, reliable information on the first two states in this channel can be obtained.\footnote{We are currently exploring the impact of Gaussian noise in the data.}  We do not present a plot of the residues but they are accurate.  In particular, the ground state residue is positive and that of the first excited state is negative, and this is obtained without any bias in the fit.

\section{Sigma terms}
The $\sigma$-term for a state $O$ is given by
\begin{equation}
\label{sigmasystem}
\sigma_{O} = m(\zeta) \frac{\partial m_O}{\partial  m(\zeta)} \,,
\end{equation}
where $m_O$ is the mass of the state, and it is a keen probe of the impact of explicit chiral symmetry breaking on a hadron's mass.  The nucleon's $\sigma$-term, $\sigma_N$, has been estimated using: chiral effective theory, e.g.\ Refs.\,\cite{Gasser:1991ce,ulf}; lattice-QCD, e.g.\ Ref.\,\cite{Leinweber:2000sa,ulf2}; and QCD-based models, e.g.\ Ref.\,\cite{lyubovitsky}.  Our recent interest in hadron $\sigma$-terms is motivated by their utility in using observational data to place constraints on the variation of nature's fundamental parameters \cite{uzan}.  In Table \ref{sigmaterms} we reproduce results calculated in Ref.\,\cite{Flambaum:2005kc} along with new results described in the text.  We note that 
\begin{equation}
\frac{\delta m_O}{m_O} = \frac{\sigma_O}{m_O} \frac{\delta m(\zeta)}{m(\zeta)} \,,
\end{equation}
so that the dimensionless quantity tabulated measures the linear relative response of the mass of interest to a fractional change in the current-quark mass, which is a renormalisation group invariant.

\begin{table}[t]
\begin{center}
\caption{\label{sigmaterms} Calculated $\sigma$-terms.  Those for $\pi$, $N$ and $\Delta$ were reported in Ref.\,\protect\cite{Flambaum:2005kc}.  The values for $\rho$ and $\omega$ listed herein were obtained via a direct analysis of the $m(\zeta)$-dependence of the vector meson mass obtained in a solution of the rainbow-ladder truncation of the quark DSE and homogeneous meson BSE.  They improve the values reported in Ref.\,\protect\cite{Flambaum:2005kc}, which were obtained using a simple fit to $m_\rho(m(\zeta))$ provided in Ref.\,\cite{marisvienna}.  All results are renormalisation-point-independent and were obtained with: $m_{u,d}(\zeta)= 3.7\,$MeV, $m_{s}(\zeta)= 82\,$MeV, $m_{c}(\zeta)= 0.97\,$GeV and $m_{b}(\zeta)= 4.1\,$GeV, where $\zeta=19\,$GeV.  Perturbative evolution can be used to determine the associated renormalisation-point-independent current-quark-masses.}

\begin{tabular*}{0.45\textwidth}{c@{\extracolsep{0ptplus1fil}}
|c@{\extracolsep{0ptplus1fil}}
 c@{\extracolsep{0ptplus1fil}}
 c@{\extracolsep{0ptplus1fil}}
 c@{\extracolsep{0ptplus1fil}}
}\hline
H & $\pi$ & $\pi_1$ & $\sigma$ &   \\
$\rule{0em}{3.5ex}\displaystyle\frac{\sigma_H}{m_H}$ 
& 0.498 & 0.017 & 0.013 &  \\\hline
H & $\rho$ & $\omega$ & $N$ & $\Delta$  \\
$\rule{0em}{3.5ex}\displaystyle\frac{\sigma_H}{m_H}$ 
& 0.021 & 0.034 & 0.064 & 0.041 \\\hline
$q$ & $u$,$d$ & $s$ & $c$ & $b$  \\
$\rule{0em}{3.5ex}\displaystyle\frac{\sigma_q}{M^E_q}$ 
& 0.023 & 0.230 & 0.637 & 0.851 \\\hline
\end{tabular*}\vspace*{-4ex}
\end{center}
\end{table}

\medskip

\hspace*{-\parindent}\underline{\textit{Radial Excitation of the Pion}}.\hspace*{0.5em}
We have a framework which enables the calculation of $\sigma_{\pi_1}$ using Eq.\,(\ref{sigmasystem}), where $\pi_1$ denotes the first pseudoscalar radial excitation.  To be specific, we can straightforwardly obtain the current-quark-mass-dependence of the $\pi_1$ in the renormalisation-group-improved rainbow-ladder (RL) DSE truncation described in Ref.\,\cite{mariscairns}.  In this truncation \mbox{$m^{\rm RL}_{\pi_1}=1.06\,$GeV \cite{Holl:2004fr}} and we find\footnote{We are currently unable to provide a reliable estimate of meson-loop corrections to the mass and $\sigma$-term of the $\pi_1$.}
\begin{equation}
\sigma_{\pi_{1}}^{\rm RL}=0.018\,\textrm{GeV}.
\end{equation}
This value is considerably smaller than that for the ground state pion: $\sigma_\pi = 0.069\,$GeV.  However, that merely serves again to emphasise the particular character of the mass of QCD's Goldstone mode and its amplification by dynamical chiral symmetry breaking.  

\medskip

\hspace*{-\parindent}\underline{\textit{Scalar Meson}}.\hspace*{0.5em}
We do not pretend to have a complete understanding of the lowest mass $0^{++}$ state in the hadron spectrum.  In this channel the rainbow-ladder truncation may be unreliable because the cancellations between higher-order terms in the systematic DSE truncation, which are so effective and important in pseudoscalar and vector channels, do not occur \cite{cdrQC2}.  This is entangled with the phenomenological difficulties encountered in understanding the scalar states below $1.4\,$GeV (see, e.g., Refs.\,\cite{MikeP}).  

Nevertheless, the rainbow-ladder DSE truncation provides a light-quark scalar-meson solution.  Combining the results from a number of sources, one finds \cite{bastirev} $m_\sigma^{\rm RL} = 0.64\pm 0.06\,$GeV.  With the kernel described in Ref.~\cite{mariscairns}, which we have used throughout, one obtains
\begin{eqnarray}
\label{massRL}
m_{\sigma}^{\rm RL} & = & 0.675\,{\rm GeV}\,,\\
\label{sigmaRL}
2\, m_\sigma^{\rm RL} \sigma_{\sigma}^{{\rm RL}} & = & (0.184\,{\rm GeV})^2 \\
& \Rightarrow & \sigma_{\sigma}^{{\rm RL}} = 0.025\,{\rm GeV}. \label{sigmaRLR}
\end{eqnarray}

The scalar meson described by the rainbow-ladder truncation has a large coupling to two pions \cite{Maris:2000ig}.  It is therefore important to consider the effect of a $\pi \pi$ loop correction to this state's mass and $\sigma$-term.  Such meson loop corrections can be estimated \cite{Leinweber:2000sa,Flambaum:2005kc,Wright:2000gg} and have a modest quantitative impact ($\lsim 15$\%) on $\sigma_\rho$, $\sigma_N$ and $\sigma_\Delta$ and a larger effect ($\lsim 30$\%) on $\sigma_\omega$.  Their effect in this case can be analysed in the same way.  

We consider a single $\pi\pi$-loop self-energy 
\begin{eqnarray}
\nonumber \lefteqn{
\Pi^{\pi\pi}_{\sigma}(m_\sigma^2)}\\
&& = \frac{-3g^2_{\sigma\pi\pi}}{16\pi^{2}} \int_{0}^{\infty}dk \frac{k^{2}\, u_{\Lambda_\sigma}(k)^2} {\omega(k)\left(\omega(k)^2-m_{\sigma}^{2}/4\right)}\,,
\label{sigmaloop}
\end{eqnarray}
where $\omega(k)=\sqrt{m_{\pi}^{2}+k^{2}}$.  The self energy both corrects $m_\sigma^{\rm RL}$ and provides for the $\sigma\to \pi\pi$ width.  In Eq.\,(\ref{sigmaloop}) 
\begin{equation}
\frac{g_{\sigma\pi\pi}}{m_\sigma^{\rm RL}}=5.51 \; \Rightarrow \; 
\Im \sqrt{s}_{\rm T} = 0.300 {\rm GeV}.
\end{equation}
This is a typical value for the imaginary part of the T-matrix pole \cite{pdg} and corresponds to 
\begin{equation}
\Gamma_{\sigma\pi\pi} = 0.55\,{\rm GeV} 
\,,
\end{equation} 
which equates numerically to $0.92\,(2\,\Im \sqrt{s}_{\rm T})$.

The function $u_{\Lambda_\sigma}(k^2) = 1/(1+k^{2}/\Lambda_{\sigma}^{2})^{2}$ in Eq.\,(\ref{sigmaloop}) is a form factor, introduced to represent the nonpointlike nature of the $\pi$ and $\sigma$ and hence the $\sigma\pi\pi$ vertex.  The analysis of Refs.\,\mbox{\cite{Bloch:1999yk,MarisFBS}} indicates that the intrinsic size of the $\sigma$ described herein is 84\% of that of the $\rho$.  We therefore choose a value of the regularisation mass-scale $\Lambda_{\sigma}=\Lambda_{\rho\pi\pi}/0.84$.  With $\Lambda_{\rho\pi\pi}=1.23\,$GeV at $m_\pi=0.14\,$GeV, as determined in Ref.\,\cite{Leinweber:2001ac}, $\Re\Pi_\sigma^{\pi\pi}((m^{\rm RL}_\sigma)^2) = - (0.395\,{\rm GeV})^2$.  Hence, the $\pi\pi$-loop acts to reduce the mass of the rainbow-ladder $\sigma$-meson and, from the shifted pole position,
\begin{equation}
\label{sigma1loop}
m_\sigma^{{\rm RL}+\pi\pi} = 0.624\,{\rm GeV.}
\end{equation}
This is an $8$\% reduction cf.\ Eq.\,(\ref{massRL}).  With all else kept fixed in Eq.\,(\ref{sigmaloop}), a variation of $\Lambda_\sigma$ by $\pm 20$\% alters the result in Eq.\,(\ref{sigma1loop}) by $\mp 6$\%.  NB.\ From Ref.\,\cite{pdg}, one might consider $0.60 \pm 0.25\,$GeV as typifying the mass of the lightest scalar meson.  

We define the $\pi\pi$-loop correction to the scalar $\sigma$-term via 
\begin{equation}
\left. m_{\pi}^{2}\frac{\partial}{\partial m_{\pi}^{2}}\, \Re\Pi_{\sigma}^{\pi\pi}\right|_{(m_{\pi}^{2})_{{\rm expt.}}}
 = -\,(0.161\,{\rm GeV})^{2}\,,
\end{equation}
and combine this with Eq.\,(\ref{sigmaRL}) accordingly:
\begin{eqnarray}
\nonumber \lefteqn{m(\zeta) \frac{\partial\,\Re m_\sigma^2 }{\partial m(\zeta)}
=(0.090\,{\rm GeV})^2}\\
%
& &=:  2 \,m_\sigma^{\Re({\rm 1-loop})} \, \sigma_\sigma^{\Re({\rm 1-loop})}\\
&  \Rightarrow & \sigma_\sigma^{\Re({\rm 1-loop})} = 0.0073 \,{\rm GeV}\,,
\label{sigmaresult}
\end{eqnarray}
since $m_\sigma^{\Re({\rm 1-loop})}=0.547\,$GeV.  Equation (\ref{sigmaresult}) is a $71$\% reduction cf.\ Eq.\,(\ref{sigmaRLR}).  The effect is large because the scalar-meson is very broad.  The $\Pi_\rho^{\pi\pi}$ self-energy contributes much less to properties of the rainbow-ladder $\rho$-meson because the width/mass ratio is significantly smaller \cite{Flambaum:2005kc,pichowsky}. 

It is noteworthy that a self-consistent solution of $s-(m_\sigma^{\rm RL})^2-\Pi_\sigma^{\pi\pi}(\Re s)=0$ gives a pole position 
\begin{equation}
\label{polemass}
\surd s_\sigma = 0.578- i \, 0.311\,{\rm GeV}.  
\end{equation}
The third iteration is the last to introduce a change $>1$\%, and all iterations beyond the first reduce the real part in Eq.\,(\ref{polemass}) by a total of $< 7$\%.  The effects are greater than that of the nucleon's $\pi N$ self energy \cite{Flambaum:2005kc,NpiN}.  These notes provide a gauge for the accuracy of the one loop analysis.

\medskip

\hspace*{-\parindent}\underline{\textit{Constituent quarks}}.\hspace*{0.5em}
One measure of the importance of dynamical chiral symmetry breaking to the dressed-quark mass function is the magnitude, relative to the current-quark mass, of the Euclidean constituent-quark mass; viz., 
\begin{equation}
\label{CQM}
(M^E)^2 := s  \ni s = M(s)^2 .
\end{equation}
For the current-quark-masses in Table \ref{sigmaterms}
\begin{equation}
\begin{array}{c|cccc}
Q & u,d & s & c & b \\\hline
\rule{0em}{3ex} M^E_Q\,({\rm GeV}) & 0.42 & 0.56 & 1.57 & 4.68
\end{array}
\end{equation}

A constituent-quark $\sigma$-term can subsequently be defined \cite{Flambaum:2005kc}
\begin{equation}
\label{sMEQ}
\sigma_Q := m(\zeta) \, \frac{\partial  M^E}{\partial m(\zeta)}\,.
\end{equation}
It is a renormalisation-group-invariant that can be determined from solutions of the gap equation, Eq.\,(\ref{eq:quark_prop}).  Our results are presented in Table \ref{sigmaterms}.

\begin{figure}[t]
\includegraphics[clip,width=0.45\textwidth]{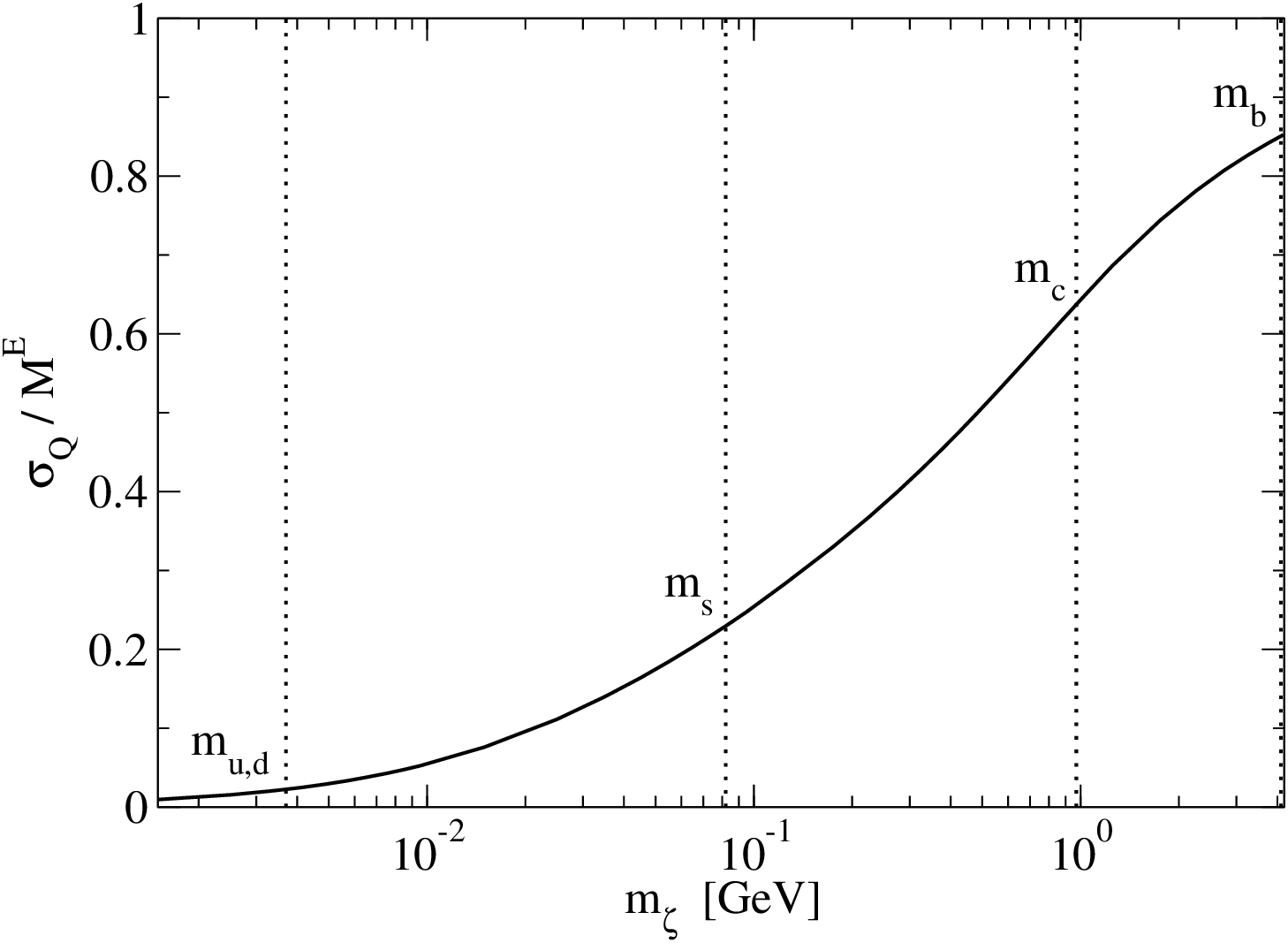}\vspace*{-4ex}

\caption{\label{fig:quark_ratio} Ratio $\sigma_Q/M^E_Q$ [Eqs.\,(\protect\ref{CQM}) \& (\protect\ref{sMEQ})].
It is a measure of the current-quark-mass-dependence of dynamical chiral symmetry breaking.  The vertical dotted lines correspond to the $u=d$, $s$, $c$ and $b$ current-quark masses listed in Table \protect\ref{sigmaterms}.\vspace*{-4ex}}
\end{figure}

In Fig.\,\ref{fig:quark_ratio} we depict $\sigma_{Q}/M^E_Q$.  It is a measure of the effect on the dressed-quark mass-function of explicit chiral symmetry breaking compared with the sum of the effects of explicit and dynamical chiral symmetry breaking.  One anticipates that for light-quarks this ratio must vanish because the magnitude of their constituent-mass owes primarily to dynamical chiral symmetry breaking, while for heavy-quarks it should approach one.  The figure confirms these expectations.

\section{Summary and Conclusion}

We presented arguments which indicate that using solely spacelike information about a Schwinger function, it may only be possible to extract properties of bound states with masses $\lsim 1.2\,$GeV.  With perfect spacelike information it is possible to accurately reconstruct the pole contributions from states which satisfy this bound.  However, such reconstruction methods provide no information beyond that which is already available in studies that employ physical (timelike) bound state momenta.  We illustrated this via the Dyson-Schwinger equations.  

In the context of numerical simulations of lattice-regularised QCD, we speculate that it may be possible to reach higher mass states by employing lattice data to constrain the infrared behaviour of DSE integral equation kernels and subsequently using the DSEs to provide information on Schwinger functions at timelike momenta.

The $\sigma$ term is one useful gauge of the impact of explicit chiral symmetry breaking on a hadron's mass.  Such information is important for the interpretation of measurements that indicate a spacial and/or temporal variation in Nature's fundamental parameters.  We calculated and reported $\sigma$-terms for a range of hadrons: the ground state pion; the pion's first radial excitation; a light scalar meson; the $\rho$ and $\omega$; the nucleon and $\Delta$; and the $u$, $s$, $c$ and $b$ constituent-quarks.  


The analysis of the scalar channel is interesting.  It is consistent with a picture of the lightest scalar as a bound state of a dressed-quark and \mbox{-antiquark} combined with a considerable two-pion component, which reduces the quark-core mass by $\lsim 15$\% but dramatically alter the current-quark-mass-dependence of the $0^{++}$ pole position.

The constituent-quark $\sigma$-terms provide a tool that is useful for assessing the flavour-dependence of dynamical chiral symmetry breaking.

\medskip

\hspace*{-\parindent}\textbf{Acknowledgments.}\hspace*{0.5em}
We avow discussions with P.\,O.~Bowman, V.\,V.~Flambaum, A.~Krass\-nigg, D.\,B.~Leinweber, P.\,C.~Tandy and A.\,G.~Williams.
This work was supported by: 
Dept.\ of Energy, Office of Nucl.\ Phys., contract nos.\ DE-FG02-00ER41135 and W-31-109-ENG-38; 
National Science Foundation grant no.\ PHY-0301190;
%
%
and the \textit{A.\,v.\ Humboldt-Stiftung} via a \textit{F.\,W.\ Bessel Forschungspreis}.
It benefited from the facilities of ANL's Computing Resource Center.

\begingroup\raggedright

\endgroup


\begin{thebibliography}{10}

\bibitem{Maris:1997hd}
P.~Maris, C.\,D.~Roberts and P.~C. Tandy, {\em Phys. Lett.} {\bf B\,420} (1998)
  pp.\, 267-273,
\href{http://www.arXiv.org/abs/nucl-th/9707003}{{\tt nucl-th/9707003}}.

\bibitem{Maris:1997tm}
P.~Maris and C.\,D.~Roberts, {\em Phys. Rev.} {\bf C\,56} (1997) pp.\, 3369-3383,
\href{http://www.arXiv.org/abs/nucl-th/9708029}{{\tt nucl-th/9708029}}.

\bibitem{bastirev} C.\,D.\ Roberts and S.\,M.\ Schmidt,
\emph{Prog.\ Part.\ Nucl.\ Phys}.\  {\bf 45} (2000) pp.\, S1-S103,
\href{http://www.arXiv.org/abs/nucl-th/0005064}{{\tt nucl-th/0005064}}.

\bibitem{reinhardrev} R.~Alkofer and L.~von~Smekal, 
\emph{Phys.\ Rept}.\ {\bf 353} (2001) pp.\ 281-465,
\href{http://www.arXiv.org/abs/hep-ph/0007355}{{\tt hep-ph/0007355}}.

\bibitem{Maris:2003vk} P.~Maris and C.\,D.~Roberts,
\emph{Int.\ J.\ Mod.\ Phys}.\ {\bf E\,12} (2003) pp.\ 297-365, 
\href{http://www.arXiv.org/abs/nucl-th/0301049}{{\tt nucl-th/0301049}}.

\bibitem{munczek} H.\,J.~Munczek,
  \emph{Phys.\ Rev}.\ \textbf{D\,52} (1995) pp.\,4736-4740,
\href{http://www.arXiv.org/abs/hep-ph/9411239}{{\tt hep-ph/9411239}}.
  
\bibitem{bender} A.~Bender, C.\,D.~Roberts and L.~von Smekal, 
\emph{Phys.\ Lett}.\ {\bf B\,380} (1997) pp.~7-12,
\href{http://www.arXiv.org/abs/nucl-th/9602012}{{\tt nucl-th/9602012}}.

\bibitem{mandarvertex} M.\,S.~Bhagwat, A.~H\"oll, A.~Krassnigg, C.\,D.~Roberts and P.\,C.~Tandy, \emph{Phys.\ Rev}.\ \textbf{C\,70} (2004) 035205 (15 pages),
\href{http://www.arXiv.org/abs/nucl-th/0403012}{{\tt nucl-th/0403012}}.

\bibitem{mariscairns} P.~Maris and P.\,C.~Tandy, ``QCD modeling of hadron physics,'' in these proceedings, 
\href{http://www.arXiv.org/abs/nucl-th/0511017}{{\tt nucl-th/0511017}}.

\bibitem{Holl:2004fr}
A.~H\"oll, A.~Krassnigg and C.\,D.~Roberts, {\em Phys. Rev.} {\bf C\,70} (2004) 042203(R) (5 pages),
\href{http://www.arXiv.org/abs/nucl-th/0406030}{{\tt nucl-th/0406030}}.

\bibitem{latticecorrelator}
D.\,B.~Leinweber, W.~Melnitchouk, D.\,G.~Richards, A.\,G.~Williams and J.\,M.~Zanotti, {\em Lect. Notes Phys.} \textbf{663} (2005) pp.\,71-112,
\href{http://www.arXiv.org/abs/nucl-th/0406032}{{\tt nucl-th/0406032}}.

\bibitem{Gasser:1991ce}
J.~Gasser, H.~Leutwyler, and M.~E. Sainio, {\em Phys. Lett.} {\bf B\,253} (1991)
pp.\,252-259.

\bibitem{ulf} M.~Frink, U.-G.~Meissner and I.~Scheller,
  \emph{Eur.\ Phys.\ J}.\ \textbf{A\,24} (2005) pp.\,395-409,
\href{http://www.arXiv.org/abs/hep-lat/0501024}{{\tt hep-lat/0501024}}.

\bibitem{Leinweber:2000sa}
D.~B. Leinweber, A.~W. Thomas and S.~V. Wright, {\em Phys. Lett.} {\bf B\,482}
  (2000) pp.\,109-113,
\href{http://www.arXiv.org/abs/hep-lat/0001007}{{\tt hep-lat/0001007}}.

\bibitem{ulf2} V.~Bernard, T.\,R.~Hemmert and U.-G.~Meissner,
  \emph{Phys.\ Lett}.\ \textbf{B\,622} (2005) pp.\,141-150,
\href{http://www.arXiv.org/abs/hep-lat/0503022}{{\tt hep-lat/0503022}}.

\bibitem{lyubovitsky} V.\,E.~Lyubovitskij, T.~Gutsche, A.~Faessler and E.\,G.~Drukarev,
  \emph{Phys.\ Rev}.\ \textbf{D\,63} (2001) 054026 (9 pages),
\href{http://www.arXiv.org/abs/hep-ph/0009341}{{\tt hep-ph/0009341}}.

\bibitem{uzan} J.\,P.~Uzan,
  \emph{Rev.\ Mod.\ Phys}.\  {\bf 75} (2003) pp.\,403-456,
\href{http://www.arXiv.org/abs/hep-ph/0205340}{{\tt hep-ph/0205340}}.

\bibitem{Flambaum:2005kc}
V.\,V.~Flambaum, A.~H\"oll, P.~Jaikumar, C.\,D.~Roberts, and S.\,V.~Wright, ``Sigma
  Terms of Light-Quark Hadrons,'' to appear in \emph{Few Body Systems}, 
\href{http://www.arXiv.org/abs/nucl-th/0510075}{{\tt nucl-th/0510075}}.

\bibitem{marisvienna} P.~Maris, ``Continuum QCD and light mesons,'' 
in \emph{Proc.\ of the Int.\ Conf.\ on Quark Confinement and the Hadron Spectrum IV}, eds.\ W.~Lucha and Kh.~Maung Maung (World Scientific, Singapore, 1997) pp.\,163-175,
\href{http://www.arXiv.org/abs/nucl-th/0009064}{{\tt nucl-th/0009064}}.

\bibitem{cdrQC2} C.\,D.~Roberts, ``Confinement, diquarks and Goldstone's theorem,'' in \emph{Proc.\ of the Int.\ Conf.\ on Quark Confinement and the Hadron Spectrum II}, eds.\ N.~Brambilla and G.\,M.~Prosperi (World Scientific, Singapore, 1997) pp.\,224-230,
\href{http://www.arXiv.org/abs/nucl-th/9609039}{{\tt nucl-th/9609039}}.

\bibitem{MikeP} M.\,R.~Pennington, ``Low Energy Hadron Physics,'' in  \emph{Frascati 1999, Physics and detectors for DAPHNE}, pp.\,43-58, 
\href{http://www.arXiv.org/abs/hep-ph/0001183}{{\tt hep-ph/0001183}};
M.~R.~Pennington, ``Scalars in the hadron world: The Higgs sector of the strong interaction,''
\href{http://www.arXiv.org/abs/hep-ph/0509265}{{\tt hep-ph/0509265}}.

\bibitem{Maris:2000ig}
  P.~Maris, C.\,D.~Roberts, S.\,M.~Schmidt and P.\,C.~Tandy,
  \emph{Phys.\ Rev}.\ \textbf{C\,63} (2001) 025202 (12 pages)
\href{http://www.arXiv.org/abs/nucl-th/0001064}{{\tt nucl-th/0001064}}.

\bibitem{Wright:2000gg}
S.~V. Wright, D.~B. Leinweber, and A.~W.~Thomas, {\em Nucl. Phys.} {\bf A\,680}
  (2000) pp.\,137-140,
\href{http://www.arXiv.org/abs/nucl-th/0005003}{{\tt nucl-th/0005003}}.

\bibitem{pdg} S.~Eidelman \textit{et al.}  
\href{http://www.slac.stanford.edu/spires/find/hep/www?rawcmd=FIND+key+5932076}%
{[Particle Data Group]},
  \emph{Phys.\ Lett}.\ \textbf{B\,592} (2004) 1.

\bibitem{Bloch:1999yk}
J.\,C.\,R.~Bloch, M.\,A.~Ivanov, T.~Mizutani, C.\,D.~Roberts, and S.\,M.~Schmidt,
  {\em Phys. Rev.} {\bf C\,62} (2000) 025206,
\href{http://www.arXiv.org/abs/nucl-th/9910029}{{\tt nucl-th/9910029}}.

\bibitem{MarisFBS} P.~Maris,
  \emph{Few Body Syst}.\  {\bf 32}, 41 (2002),
\href{http://www.arXiv.org/abs/nucl-th/0204020}{{\tt nucl-th/0204020}}.

\bibitem{Leinweber:2001ac}
D.\,B.~Leinweber, A.\,W.~Thomas, K.~Tsushima, and S.\,V.~Wright, {\em Phys. Rev.} \textbf{D\,64} (2001) 094502 (8 pages),
\href{http://www.arXiv.org/abs/hep-lat/0104013}{{\tt hep-lat/0104013}}.

\bibitem{pichowsky} M.\,A.~Pichowsky, S.~Walawalkar and S.~Capstick,
  \emph{Phys.\ Rev}.\ \textbf{D\,60} (1999) 054030 (21 pages),
\href{http://www.arXiv.org/abs/nucl-th/9904079}{{\tt nucl-th/9904079}}.
  
\bibitem{NpiN} M.\,B.~Hecht, M.~Oettel, C.\,D.~Roberts, S.\,M.~Schmidt, P.\,C.~Tandy and A.\,W.~Thomas,
  \emph{Phys.\ Rev}.\ \textbf{C\,65} (2002) 055204,
\href{http://www.arXiv.org/abs/nucl-th/0201084}{{\tt nucl-th/0201084}}.

\end{thebibliography}
\end{document}